\documentclass{jpsj3}
\usepackage{txfonts}
\usepackage{bm}
\usepackage{graphicx}
\usepackage{color}
\usepackage{float}

\title{Kernel Learning for Regression via Quantum Annealing-Based Spectral Sampling}

\author{Yasushi Hasegawa$^{1,2}$\thanks{hasegawa.yasushi.s6@dc.tohoku.ac.jp}, and Masayuki Ohzeki$^{1,3,4,5}$}
\inst{$^1$Graduate School of Information Sciences, Tohoku University, Sendai 980-8579, Japan \\
$^2$Technology Strategy Division, SCSK Corporation, Koto, Tokyo 135-8110, Japan \\
$^3$Department of Physics, Institute of Science Tokyo, Meguro, Tokyo 152-8551, Japan \\
$^4$Research and Education Institute for Semiconductors and Informatics, Kumamoto University, Kumamoto 860-8555, Japan\\
$^5$Sigma-i Co., Ltd., Minato, Tokyo 108-0075, Japan }

\abst{While quantum annealing (QA) has been developed for combinatorial optimization, practical QA devices operate at finite temperature and under noise, and their outputs can be regarded as stochastic samples close to a Gibbs--Boltzmann distribution. In this study, we propose a QA-in-the-loop kernel learning framework that integrates QA not merely as a substitute for Markov-chain Monte Carlo sampling but as a component that directly determines the learned kernel for regression. Based on Bochner's theorem, a shift-invariant kernel is represented as an expectation over a spectral distribution, and random Fourier features (RFF) approximate the kernel by sampling frequencies. We model the spectral distribution with a (multi-layer) restricted Boltzmann machine (RBM), generate discrete RBM samples using QA, and map them to continuous frequencies via a Gaussian--Bernoulli transformation. Using the resulting RFF, we construct a data-adaptive kernel and perform Nadaraya--Watson (NW) regression. Because the RFF approximation based on $\cos(\bm{\omega}^{\top}\Delta\bm{x})$ can yield small negative values and cancellation across neighbors, the Nadaraya--Watson denominator $\sum_j k_{ij}$ may become close to zero. We therefore employ nonnegative squared-kernel weights $w_{ij}=k(\bm{x}_i,\bm{x}_j)^2$, which also enhances the contrast of kernel weights. The kernel parameters are trained by minimizing the leave-one-out NW mean squared error, and we additionally evaluate local linear regression with the same squared-kernel weights at inference. Experiments on multiple benchmark regression datasets demonstrate a decrease in training loss, accompanied by structural changes in the kernel matrix, and show that the learned kernel tends to improve $R^2$ and RMSE over the baseline Gaussian-kernel NW. Increasing the number of random features at inference further enhances accuracy.}

\begin{document}
\maketitle

\section{Introduction}

Regression is a fundamental task in machine learning, with broad applications in science and engineering.
A classical and widely used nonparametric approach is kernel regression, where a similarity-weighted average of observed outputs forms predictions.
In particular, the Nadaraya--Watson (NW) estimator provides a simple and interpretable mapping from a kernel-induced weight structure to the prediction \cite{nadaraya1964,watson1964}.
Kernel methods have been extensively studied as a general framework for learning with similarities
\cite{schölkopf2002,gonen2011}, and they remain attractive when one aims to control model complexity and maintain
transparency in the inference procedure.

Despite its simplicity, the NW estimator is sensitive to the choice of kernel and its hyperparameters,
and it can exhibit bias in regions with sparse data or near boundaries.
A standard remedy is to exploit local polynomial structure, e.g., local linear regression, which reduces boundary bias while keeping a kernel-weighted formulation \cite{fan2018}.
These observations motivate data-adaptive kernel construction: rather than fixing a kernel family a priori,
we seek a mechanism that can reshape the kernel to reflect dataset-dependent geometry and thereby improve predictive performance.

For shift-invariant kernels, Bochner's theorem provides a principled route to such construction by linking kernels to spectral distributions \cite{bochner1959}.
Random Fourier feature (RFF) approximate the kernel by Monte Carlo sampling of frequencies from the corresponding spectral measure, enabling scalable kernel learning without explicitly forming the full kernel matrix \cite{Rahimi2007,Liu2020}.
However, the approximation quality and downstream performance of RFF crucially depend on the chosen spectral
distribution.
This has led to research on learning spectral distributions (implicit kernel learning) to obtain flexible kernels adapted to the data \cite{Li2019,Wilson2013}.

In this work, we model the spectral distribution by restricted Boltzmann machine (RBM),
a generative model capable of representing complex probability distributions \cite{Hinton2002,Salakhutdinov2007,Upadhya2019}.
Training and sampling of RBMs typically require drawing samples from distributions close to the model equilibrium,
which is often computationally demanding.
Quantum annealing (QA) was originally developed as a heuristic for solving combinatorial optimization problems
\cite{Kadowaki1998}, but practical QA devices operate at finite temperature and are subject to noise and freezing
effects; consequently, their outputs are not restricted to the ground state and can be interpreted as stochastic
samples close to a Gibbs--Boltzmann distribution \cite{amin2015}.
This ``QA as a sampler'' perspective has motivated applications of QA to sampling-based machine learning, including RBM-related tasks, effective-temperature estimation, and hyperparameter adaptation for training RBMs on quantum annealers \cite{Benedetti2016,benedetti2017,xu2021,raymond2016,Dixit2021}. Relatedly, quantum-hardware-driven feature maps such as Quantum Kitchen Sinks have been proposed for supervised learning, indicating the potential of quantum devices as components in feature/kernel construction \cite{wilson2019,Noori2020}.

Most prior studies have positioned QA primarily as an alternative to conventional Markov-chain Monte Carlo for
sampling in RBM learning.
In contrast, our goal is to integrate QA into an end-to-end learning pipeline and to demonstrate a clear benefit on a downstream regression task.
Specifically, we propose a QA-in-the-loop kernel learning framework in which (i) an RBM parameterizes the spectral distribution of a shift-invariant kernel, (ii) QA generates samples from the RBM, (iii) the samples are transformed into frequencies and used to construct RFF, and (iv) the resulting data-adaptive kernel is employed for NW regression.
Kernel parameters are trained by minimizing the leave-one-out NW mean squared error, and we also evaluate local linear regression using the same kernel weights at inference.

The contributions of this paper are as follows.
First, we present a QA-in-the-loop kernel learning framework that constructs a data-adaptive kernel from QA-generated RBM samples via RFF.
Second, beyond viewing QA as a mere sampling substitute, we empirically show on multiple benchmark datasets that QA samples can improve the predictive accuracy of kernel regression.
Third, we analyze how the learned spectral distribution and the kernel structure evolve during training, and discuss their correspondence with performance improvements.

\section{Methods}

\subsection{Problem setting}
Let $\mathcal{D}=\{(\bm{x}_i,y_i)\}_{i=1}^{N}$ be a regression dataset, where $\bm{x}_i\in\mathbb{R}^{d}$ and
$y_i\in\mathbb{R}$.
Our objective is to learn a data-adaptive shift-invariant kernel by learning its spectral (frequency) distribution using samples obtained from a quantum annealer (QA), and to improve regression accuracy through kernel regression.

\subsection{Random Fourier features for shift-invariant kernels}
We consider a shift-invariant positive definite kernel $k(\bm{x},\bm{x}')=k(\bm{x}-\bm{x}')$.
By Bochner's theorem, such a kernel can be written as an expectation over a spectral distribution
$p_{\theta}(\bm{\omega})$ \cite{bochner1959}:
\begin{align}
k(\bm{x}-\bm{x}')
&= \mathbb{E}_{\bm{\omega}\sim p_{\theta}(\bm{\omega})}
\left[\exp\!\left(i\,\bm{\omega}^{\mathsf{T}}(\bm{x}-\bm{x}')\right)\right]
= \mathbb{E}_{\bm{\omega}\sim p_{\theta}(\bm{\omega})}
\left[\cos\!\left(\bm{\omega}^{\mathsf{T}}(\bm{x}-\bm{x}')\right)\right].
\label{eq:bochner}
\end{align}
Random Fourier feature (RFF) approximate Eq.~\eqref{eq:bochner} by Monte Carlo sampling
$\{\bm{\omega}_s\}_{s=1}^{S}\sim p_{\theta}(\bm{\omega})$ \cite{Rahimi2007}.
We use the standard cosine--sine feature map
\begin{equation}
\bm{\phi}_{\theta}(\bm{x})
= \frac{1}{\sqrt{S}}
\bigl(\cos(\bm{\omega}_1^{\mathsf{T}}\bm{x}),\ldots,\cos(\bm{\omega}_S^{\mathsf{T}}\bm{x}),
\sin(\bm{\omega}_1^{\mathsf{T}}\bm{x}),\ldots,\sin(\bm{\omega}_S^{\mathsf{T}}\bm{x})\bigr)^{\mathsf{T}},
\label{eq:rff_feature}
\end{equation}
and approximate the kernel by an inner product:
\begin{equation}
k_{\theta}(\bm{x},\bm{x}')
\approx \bm{\phi}_{\theta}(\bm{x})^{\mathsf{T}}\bm{\phi}_{\theta}(\bm{x}').
\label{eq:rff_kernel}
\end{equation}
We write $k_{ij}=k_{\theta}(\bm{x}_i,\bm{x}_j)$ and $\bm{\phi}_i=\bm{\phi}_{\theta}(\bm{x}_i)$.

\subsection{Spectral sampling by (multi-layer) RBM and quantum annealing}
To obtain a flexible $p_{\theta}(\bm{\omega})$, we model the spectral distribution with a Boltzmann-machine family model and generate discrete samples using QA.
We start from a restricted Boltzmann machine (RBM) on Ising variables
$\bm{v}\in\{-1,+1\}^{N_v}$ and $\bm{h}\in\{-1,+1\}^{N_h}$ \cite{Hinton2002,Salakhutdinov2007}:
\begin{align}
P(\bm{v},\bm{h}\mid \bm{b},\bm{c},\bm{W})
&= \frac{\exp\!\left[-E(\bm{v},\bm{h}\mid \bm{b},\bm{c},\bm{W})\right]}{Z(\bm{b},\bm{c},\bm{W})},
\label{eq:rbm_joint}\\
E(\bm{v},\bm{h}\mid \bm{b},\bm{c},\bm{W})
&= -\sum_{i} b_i v_i - \sum_{j} c_j h_j - \sum_{i,j} W_{ij} v_i h_j .
\label{eq:rbm_energy}
\end{align}
We map the RBM energy to an Ising form and use a quantum annealer as a stochastic sampler to obtain
$(\bm{v},\bm{h})$ samples.
While QA was developed as a heuristic for combinatorial optimization \cite{Kadowaki1998}, practical devices operate at finite temperature and under noise; thus their outputs can be interpreted as stochastic samples close to a Gibbs--Boltzmann distribution \cite{amin2015,Benedetti2016}.
(Embedding and scaling details are described in the experimental setting.)

\subsection{Gaussian--Bernoulli mapping to continuous frequencies}
Given a discrete sample $\bm{v}$, we generate a continuous frequency vector
$\bm{\omega}\in\mathbb{R}^{N_{\omega}}$ via a Gaussian--Bernoulli conditional model
(related to Gaussian--Bernoulli RBMs) \cite{Cho2011}:
\begin{equation}
P(\bm{\omega}\mid \bm{a},\bm{U},\bm{v},\bm{\sigma})
= \prod_{i=1}^{N_{\omega}}
\mathcal{N}\!\left(\omega_i \,\middle|\,
a_i + \sum_{j=1}^{N_v} U_{ij} v_j,\ \sigma_i^{2}\right).
\label{eq:gb_conditional}
\end{equation}
Equivalently, the corresponding conditional energy is
\begin{equation}
E(\bm{\omega}\mid \bm{a},\bm{U},\bm{v},\bm{\sigma})
= \sum_{i=1}^{N_{\omega}}
\frac{\bigl(\omega_i - (a_i+\sum_{j}U_{ij}v_j)\bigr)^2}{2\sigma_i^2}.
\label{eq:gb_energy}
\end{equation}
For numerical stability, we parameterize $\sigma_i^2=\exp(z_i)$ and optimize $z_i$.
In the experiments, we set $(N_v,N_h)=(4,4)$ and $N_{\omega}=d$.

Overall, the learnable parameters are
\begin{equation}
\theta=\{\bm{a},\bm{b},\bm{c},\bm{U},\bm{W},\bm{z}\}.
\end{equation}
A frequency sample is generated by $(\bm{v},\bm{h})\sim P(\bm{v},\bm{h}\mid\bm{b},\bm{c},\bm{W})$ (via QA),
followed by $\bm{\omega}\sim P(\bm{\omega}\mid \bm{a},\bm{U},\bm{v},\bm{\sigma})$.

\subsection{Leave-one-out Nadaraya--Watson regression with squared-kernel weights}
Using the RFF kernel approximation in Eq.~\eqref{eq:rff_kernel}, we perform Nadaraya--Watson (NW) regression
\cite{nadaraya1964,watson1964}.
In this work, we use the squared-kernel weights
\begin{equation}
w_{ij} = k_{ij}^{2}.
\label{eq:weight_square}
\end{equation}
The kernel value $k_{ij}$ is approximated by random Fourier features as an empirical average of $\cos(\bm{\omega}^{\top}(\bm{x}_i-\bm{x}_j))$.
With a finite number of random features, this approximation can take small negative values and may suffer from cancellation across neighbors, so the NW normalizer $\sum_{j\neq i} k_{ij}$ (or its RFF approximation) can become close to zero for some $i$, leading to numerical instability.
To avoid such near-zero denominators while keeping the same similarity structure, we adopt squared-kernel weights, which are nonnegative and less prone to cancellation.
To mitigate overfitting during kernel learning, we employ leave-one-out (LOO) NW prediction on the training set:
\begin{equation}
\hat{y}_i^{(-i)}
= \frac{\sum_{j\neq i} w_{ij} y_j}{\sum_{j\neq i} w_{ij}} .
\label{eq:nw_loo}
\end{equation}
In our implementation, we additionally add a small constant $\varepsilon>0$ to the denominator in Eq.~\ref{eq:nw_loo} for robustness.

\subsection{Loss function}
We minimize the mean squared error (MSE) based on LOO predictions:
\begin{equation}
L(\theta)
= \frac{1}{N}\sum_{i=1}^{N}\bigl(y_i-\hat{y}_i^{(-i)}\bigr)^2.
\label{eq:mse_loo}
\end{equation}
Since $\hat{y}_i^{(-i)}$ depends on $\theta$ through $w_{ij}$, $k_{ij}$, and ultimately the spectral distribution
$p_{\theta}(\bm{\omega})$, the kernel is learned end-to-end from the regression error.

\subsection{Gradient of the kernel-learning objective}
\label{sec:gradient}
Let
$A_i=\sum_{j\neq i} w_{ij}y_j$ and $B_i=\sum_{j\neq i} w_{ij}$ so that $\hat{y}_i^{(-i)}=A_i/B_i$.
Differentiating Eq.~\eqref{eq:nw_loo} yields
\begin{equation}
\frac{\partial \hat{y}_i^{(-i)}}{\partial \theta}
= \frac{1}{B_i}\sum_{j\neq i}\bigl(y_j-\hat{y}_i^{(-i)}\bigr)\frac{\partial w_{ij}}{\partial \theta}.
\label{eq:dyhat}
\end{equation}
From Eq.~\eqref{eq:mse_loo}, we obtain
\begin{equation}
\frac{\partial L}{\partial \theta}
= -\frac{2}{N}\sum_{i=1}^{N}\bigl(y_i-\hat{y}_i^{(-i)}\bigr)
\frac{\partial \hat{y}_i^{(-i)}}{\partial \theta}.
\label{eq:dL_general}
\end{equation}
Because $w_{ij}=k_{ij}^{2}$ (Eq.~\eqref{eq:weight_square}),
\begin{equation}
\frac{\partial w_{ij}}{\partial \theta} = 2k_{ij}\frac{\partial k_{ij}}{\partial \theta}.
\label{eq:dw}
\end{equation}

Next, the kernel is written as
\begin{equation}
k(\bm{x}_i,\bm{x}_j)
= \mathbb{E}_{\bm{\omega}\sim p_{\theta}(\bm{\omega})}
\left[\cos\!\left(\bm{\omega}^{\mathsf{T}}(\bm{x}_i-\bm{x}_j)\right)\right],
\label{eq:k_expect}
\end{equation}
and is approximated by RFF sampling. Using the score-function identity,
\begin{equation}
\frac{\partial k_{ij}}{\partial \theta}
= \mathbb{E}_{\bm{\omega}\sim p_{\theta}}
\left[
\cos\!\left(\bm{\omega}^{\mathsf{T}}(\bm{x}_i-\bm{x}_j)\right)
\frac{\partial \log p_{\theta}(\bm{\omega})}{\partial \theta}
\right]
\approx
\frac{1}{S}\sum_{s=1}^{S}
\cos\!\left(\bm{\omega}_s^{\mathsf{T}}(\bm{x}_i-\bm{x}_j)\right)
\frac{\partial \log p_{\theta}(\bm{\omega}_s)}{\partial \theta}.
\label{eq:dk_score}
\end{equation}
When $p_{\theta}$ is represented in a Boltzmann form $p_{\theta}(\bm{\omega})=\exp[-E_{\theta}(\bm{\omega})]/Z_{\theta}$,
\begin{equation}
\frac{\partial \log p_{\theta}(\bm{\omega})}{\partial \theta}
=
-\frac{\partial E_{\theta}(\bm{\omega})}{\partial \theta}
-\frac{1}{Z(\theta)}\frac{\partial Z(\theta)}{\partial \theta},
\label{eq:score_boltzmann}
\end{equation}
and the second (model-expectation) term can be approximated using the same set of samples.
Substituting Eqs.~\eqref{eq:dyhat}--\eqref{eq:score_boltzmann} gives the gradient used to update $\theta$.

\subsection{Local linear regression with squared-kernel weights (inference-only)}
NW regression can suffer from boundary bias; local linear regression (LLR) is a classical remedy that reduces such bias while keeping a kernel-weighted form \cite{fan2018}.
In this study, LLR is used only at inference with the same squared-kernel weights.

For a query point $\bm{x}_{*}$, define $w_{*j}=k_{\theta}(\bm{x}_{*},\bm{x}_j)^{2}$ and assume a local linear model
$y \approx \beta_0 + \bm{\beta}^{\mathsf{T}}(\bm{x}-\bm{x}_{*})$.
Let
\begin{equation}
\bm{X}_{*}=
\begin{pmatrix}
1 & (\bm{x}_1-\bm{x}_{*})^{\mathsf{T}}\\
\vdots & \vdots\\
1 & (\bm{x}_N-\bm{x}_{*})^{\mathsf{T}}
\end{pmatrix},
\qquad
\bm{y}=(y_1,\ldots,y_N)^{\mathsf{T}},
\qquad
\bm{W}_{*}=\mathrm{diag}(w_{*1},\ldots,w_{*N}),
\end{equation}
and solve the weighted least squares problem
\begin{equation}
\hat{\bm{\beta}}_{*}
= \arg\min_{\bm{\beta}\in\mathbb{R}^{d+1}}
\left\|\bm{W}_{*}^{1/2}\bigl(\bm{y}-\bm{X}_{*}\bm{\beta}\bigr)\right\|_{2}^{2}.
\end{equation}
The LLR prediction is given by the intercept term $\hat{y}_{\mathrm{LLR}}(\bm{x}_{*})=\hat{\beta}_{*,0}$.

We use LLR as an endpoint correction at inference time.
Here, an endpoint query is defined as a test input $\bm{x}_*$ such that at least one of its coordinates lies in the lower or upper $1\%$ of the corresponding training marginal distribution, i.e.,
\begin{equation}
\exists m \in \{1,\dots,d\} \ \text{s.t.}\ 
x_{*,m} \le q^{\mathrm{train}}_{m}(0.01)
\ \ \text{or}\ \ 
x_{*,m} \ge q^{\mathrm{train}}_{m}(0.99),
\end{equation}
where $q^{\mathrm{train}}_{m}(\alpha)$ denotes the $\alpha$-quantile of the $m$-th feature over the training set.
For such endpoint queries, we replace the NW prediction with the LLR prediction computed with the same squared-kernel weights, i.e., we output $\hat{y}(\bm{x}_*)=\hat{y}_{\mathrm{LLR}}(\bm{x}_*)$ instead of the NW estimate.
Otherwise, we use the standard NW prediction.

\section{Results}

\subsection{Datasets and experimental setup}
We evaluated the proposed kernel learning for regression on the following benchmark datasets.
Each dataset was split into training and test sets with an 8:2 ratio, and all input features were standardized.

\begin{itemize}
  \item \textit{bodyfat}: a regression dataset for estimating body fat percentage from body measurements ($d=14$, $N=252$).
  \item \textit{Mackey--Glass (mg)}: a time-series dataset generated from the Mackey--Glass equation ($d=6$, $N=700$).
  \item \textit{energy efficiency}: a regression dataset for predicting building energy loads from 8 variables
  (relative compactness, surface area, wall area, roof area, overall height, orientation, glazing area, and glazing area distribution)
  ($d=8$, $N=768$).
  \item \textit{Concrete Compressive Strength (ccs)}: a regression dataset for predicting compressive strength from mixture proportions and age
  ($d=8$, $N=700$).
\end{itemize}

As evaluation metrics, we used the coefficient of determination ($R^2$) and the root mean squared error (RMSE).
The coefficient of determination is defined as
\begin{equation}
  R^2 = 1 - \frac{\sum_i (y_i - \hat{y}_i)^2}{\sum_i (y_i - \bar{y})^2},
  \qquad
  \bar{y} = \frac{1}{N}\sum_i y_i .
\end{equation}

As baselines, we used (i) Nadaraya--Watson (NW) regression with a Gaussian kernel and
(ii) Support Vector Regression (SVR) with a Gaussian kernel.
Hyperparameters for NW and SVR were tuned with Optuna \cite{Akiba2019}, and SVR tuning employed the hold-out method.

We utilized the D-Wave Advantage system (Solver: Advantage\_system4.1) as a stochastic sampler for generating RBM states $(\bm{v},\bm{h})$.
At each training iteration, we submitted an Ising problem corresponding to Eq.~(5) and requested $R$ reads with an annealing time of $1~\mu\mathrm{s}$, where we set $R = N_{\mathrm{train}}/2$.
The returned samples were then used to generate frequency vectors $\bm{\omega}$ via the Gaussian--Bernoulli mapping (Sec.~2.4).

\subsection{Training dynamics and kernel adaptation}
\begin{figure}[t]
  \centering
  \includegraphics[width=0.95\linewidth]{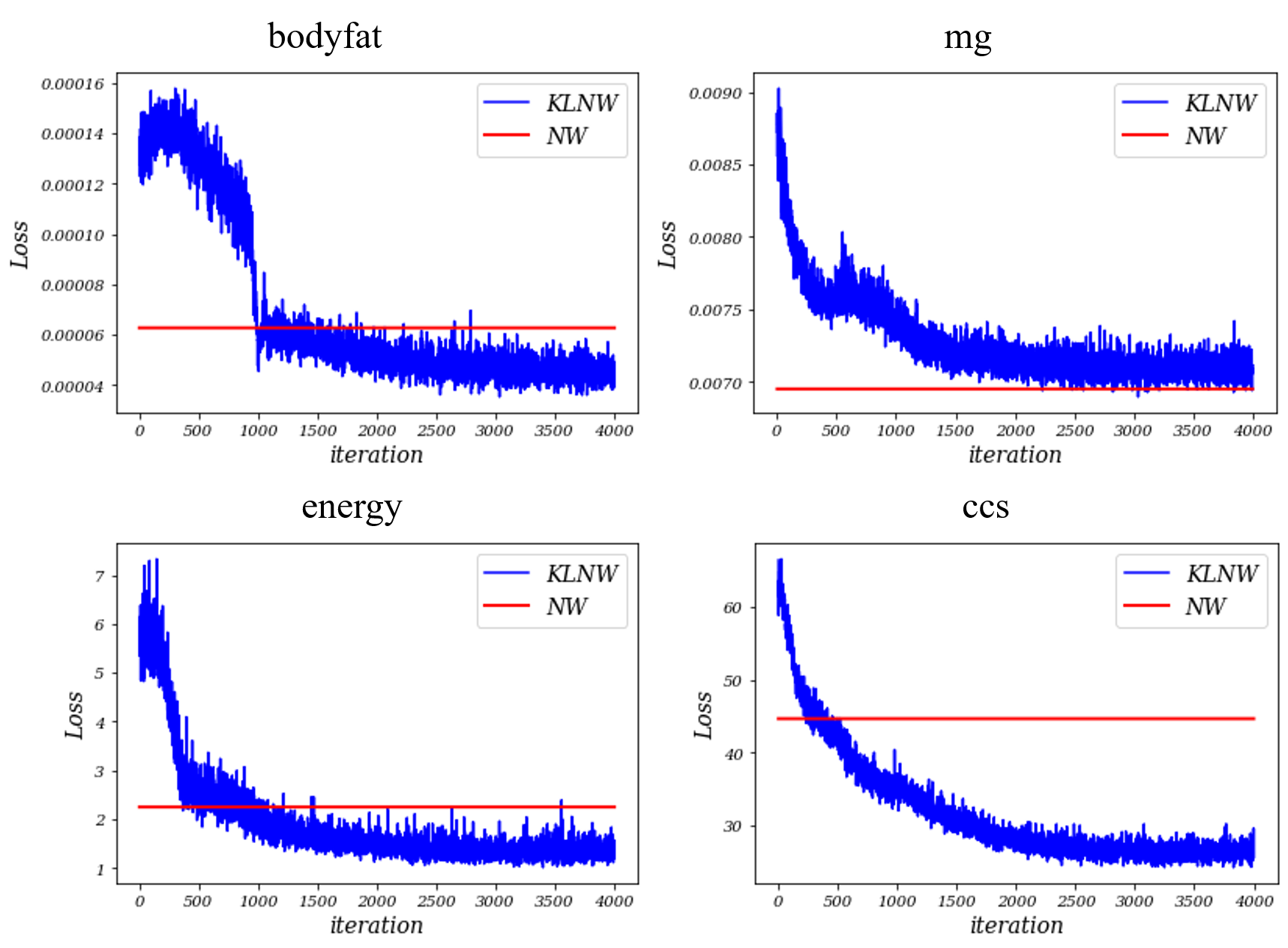}
  \caption{Training loss trajectory and comparison with the NW loss.
  The loss decreases as training proceeds.}
  \label{fig:loss}
\end{figure}

\begin{figure}[t]
  \centering
  \includegraphics[width=0.95\linewidth]{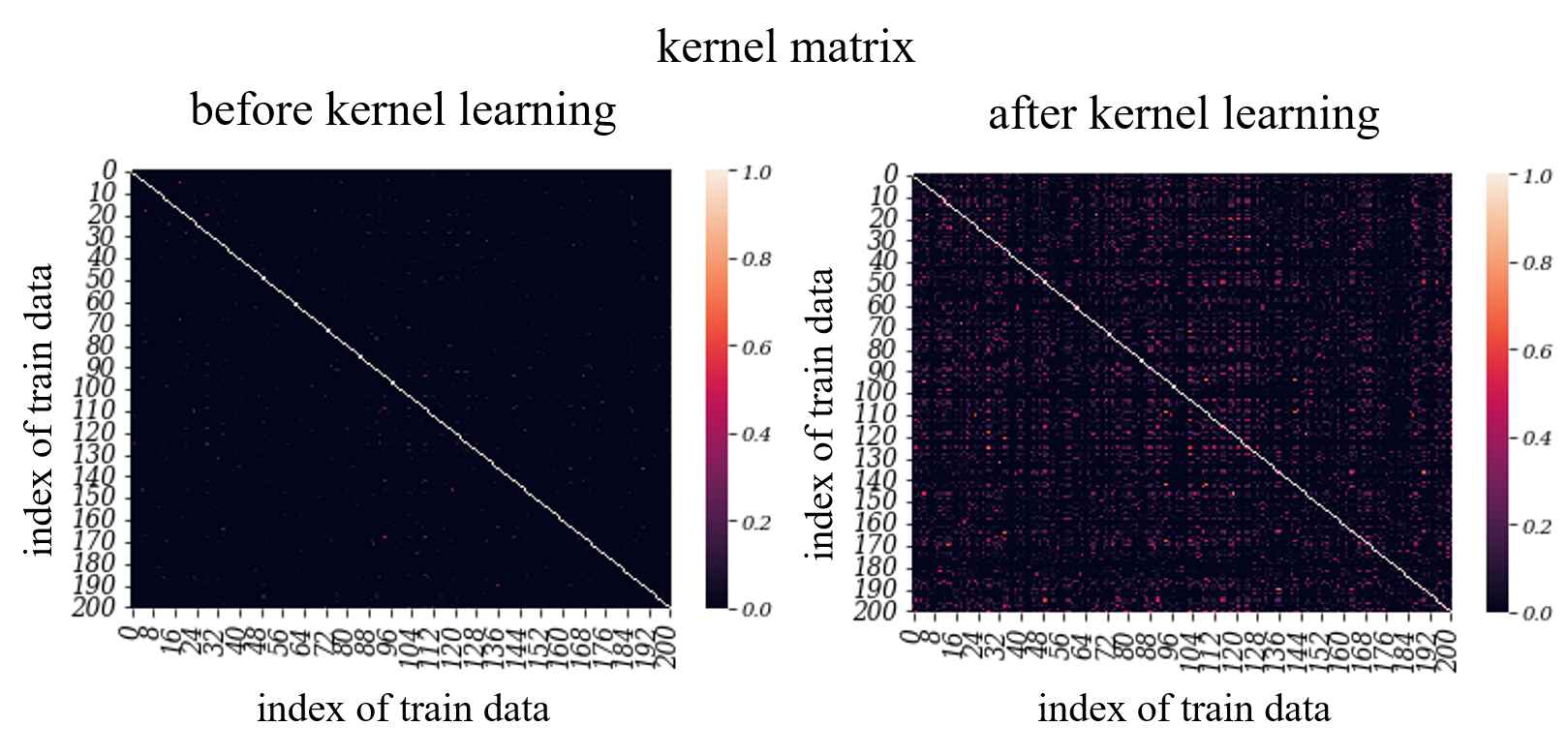}
  \caption{Kernel matrix on the \textit{bodyfat} dataset.
  Training changes off-diagonal kernel components $k_{ij}$, indicating kernel adaptation.}
  \label{fig:kernel}
\end{figure}

Figure~\ref{fig:loss} shows that the regression loss decreases as training proceeds, indicating that the learned kernel improves the leave-one-out NW objective.
Figure~\ref{fig:kernel} visualizes the kernel matrix on the \textit{bodyfat} dataset; notably, off-diagonal kernel entries change after training, demonstrating that the kernel structure is adapted to data.

\subsection{Effect of the number of random features at inference}
\begin{figure}[t]
  \centering
  \includegraphics[width=0.95\linewidth]{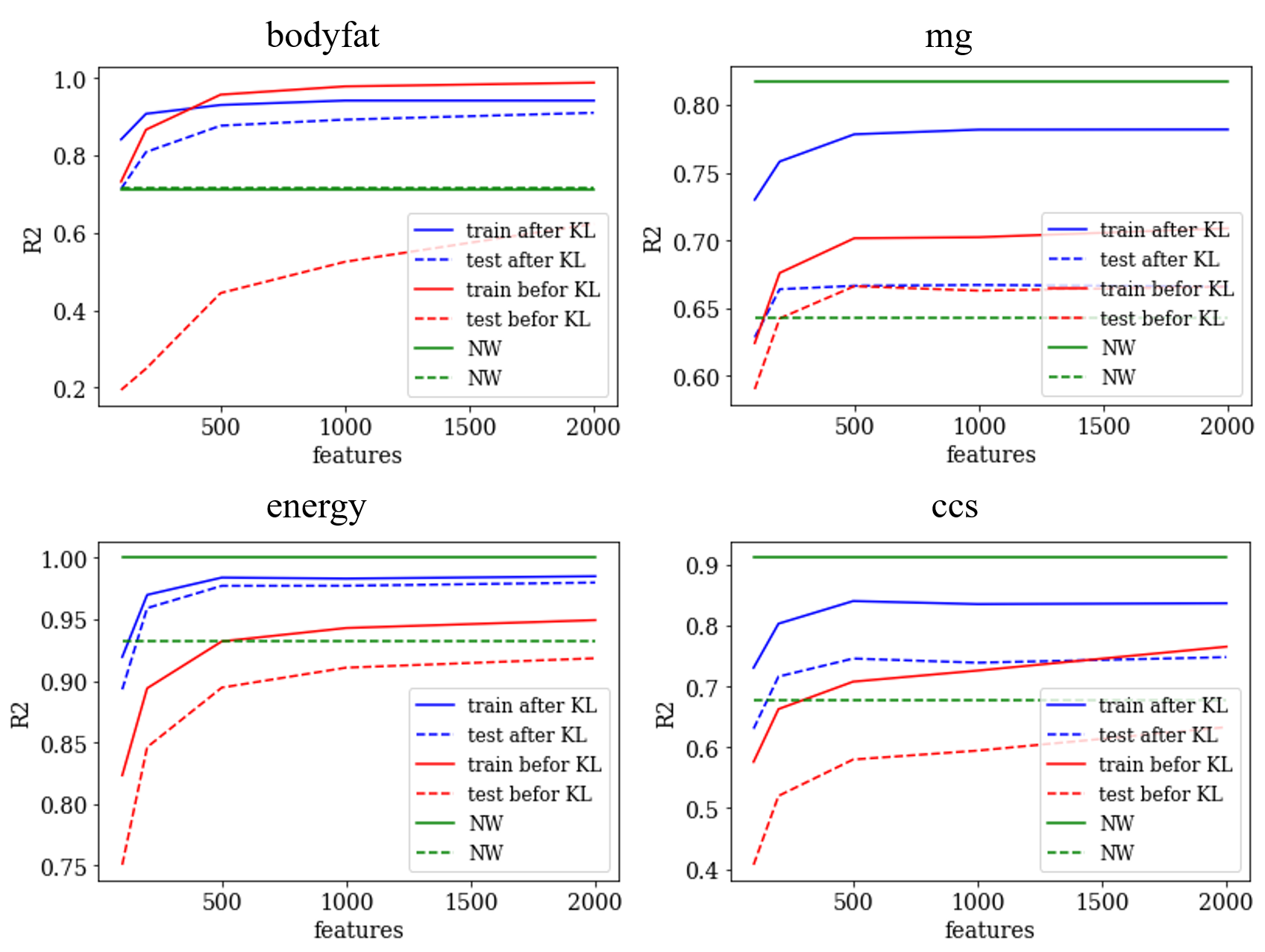}
  \caption{Comparison of $R^2$ versus the number of random features $S$ at inference.
  Red: before training, blue: after training, green: NW baseline.
  Solid lines: training, dashed lines: test.}
  \label{fig:r2}
\end{figure}

\begin{figure}[t]
  \centering
  \includegraphics[width=0.95\linewidth]{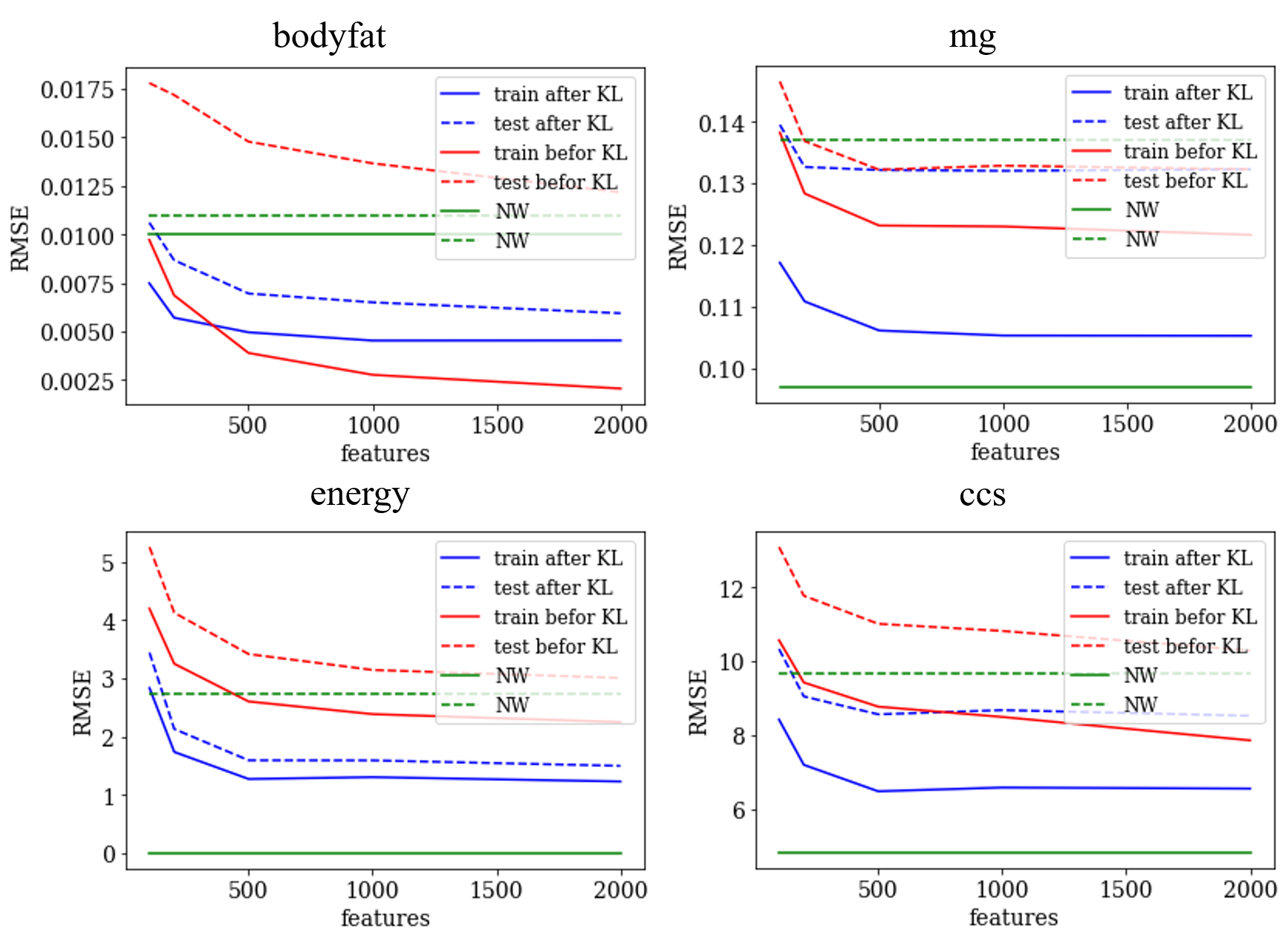}
  \caption{Comparison of RMSE versus the number of random features $S$ at inference.
  Red: before training, blue: after training, green: NW baseline.
  Solid lines: training, dashed lines: test.}
  \label{fig:rmse}
\end{figure}

We also investigated the impact of the number of random features $S$ at inference.
Figure~\ref{fig:r2} compares $R^2$ as a function of $S$, and Fig.~\ref{fig:rmse} compares RMSE as a function of $S$.
In both metrics, using the learned kernel after training yields better performance than before training.
Moreover, increasing $S$ at inference improves accuracy, suggesting that the approximation error due to finite random features can be reduced on the inference side.

\subsection{Quantitative comparison across datasets (NW regression)}
Tables~\ref{tab:nw_r2} and \ref{tab:nw_rmse} summarize $R^2$ and RMSE of NW regression using the learned kernel
(KLNW) for different inference feature sizes $S \in \{100,200,500,1000,2000\}$.
For reference, we also report the Gaussian-kernel NW baseline (NW) in the rightmost columns.
Overall, KLNW tends to improve both $R^2$ and RMSE compared with the NW baseline, with clear gains on \textit{bodyfat} and \textit{ccs}.

\begin{table*}[t]
  \centering
  \caption{$R^2$ of NW regression with the learned kernel (KLNW). Each cell reports (Train, Test).}
  \label{tab:nw_r2}
  \scriptsize
  \begin{tabular}{lcccccccccccc}
    \hline
    & \multicolumn{2}{c}{$S{=}100$} & \multicolumn{2}{c}{$S{=}200$} & \multicolumn{2}{c}{$S{=}500$}
    & \multicolumn{2}{c}{$S{=}1000$} & \multicolumn{2}{c}{$S{=}2000$} & \multicolumn{2}{c}{NW} \\
    Dataset & Train & Test & Train & Test & Train & Test & Train & Test & Train & Test & Train & Test \\
    \hline
    bodyfat & 0.841 & 0.713 & 0.907 & \textbf{0.808} & 0.930 & \textbf{0.877} & 0.942 & \textbf{0.892} & 0.941 & \textbf{0.910} & 0.711 & 0.715 \\
    mg      & 0.730 & 0.623 & 0.758 & \textbf{0.664} & 0.778 & \textbf{0.667} & 0.782 & \textbf{0.667} & 0.781 & \textbf{0.666} & 0.817 & 0.643 \\
    energy  & 0.919 & 0.893 & 0.970 & \textbf{0.959} & 0.984 & \textbf{0.977} & 0.983 & \textbf{0.977} & 0.985 & \textbf{0.980} & 1.000 & 0.932 \\
    ccs     & 0.731 & 0.630 & 0.803 & \textbf{0.716} & 0.840 & \textbf{0.746} & 0.835 & \textbf{0.739} & 0.837 & \textbf{0.748} & 0.912 & 0.677 \\
    \hline
  \end{tabular}
\end{table*}

\begin{table*}[t]
  \centering
  \caption{RMSE of NW regression with the learned kernel (KLNW). Each cell reports (Train, Test).}
  \label{tab:nw_rmse}
  \scriptsize
  \begin{tabular}{lcccccccccccc}
    \hline
    & \multicolumn{2}{c}{$S{=}100$} & \multicolumn{2}{c}{$S{=}200$} & \multicolumn{2}{c}{$S{=}500$}
    & \multicolumn{2}{c}{$S{=}1000$} & \multicolumn{2}{c}{$S{=}2000$} & \multicolumn{2}{c}{NW} \\
    Dataset & Train & Test & Train & Test & Train & Test & Train & Test & Train & Test & Train & Test \\
    \hline
    bodyfat & 0.007 & 0.011 & 0.006 & \textbf{0.009} & 0.005 & \textbf{0.007} & 0.005 & \textbf{0.007} & 0.005 & \textbf{0.006} & 0.010 & 0.011 \\
    mg      & 0.117 & 0.139 & 0.111 & \textbf{0.133} & 0.106 & \textbf{0.132} & 0.105 & \textbf{0.132} & 0.105 & \textbf{0.132} & 0.097 & 0.137 \\
    energy  & 2.832 & 3.438 & 1.735 & \textbf{2.128} & 1.272 & \textbf{1.592} & 1.303 & \textbf{1.590} & 1.229 & \textbf{1.497} & 0.000 & 2.732 \\
    ccs     & 8.424 & 10.331 & 7.200 & \textbf{9.051} & 6.485 & \textbf{8.566} & 6.587 & \textbf{8.680} & 6.558 & \textbf{8.527} & 4.828 & 9.653 \\
    \hline
  \end{tabular}
\end{table*}

\subsection{Local linear regression at inference}
Although NW regression is simple, it may suffer from bias near boundaries or under uneven sample densities because it does not explicitly exploit local linear structure.
Therefore, we also evaluated local linear regression (LLR) at inference time, using the same kernel-based weights as NW. Following Sec.~2.8, we apply LLR only to endpoint queries (defined by the 1\%/99\% training quantiles) and use the standard NW prediction otherwise.
Tables~\ref{tab:llr_r2} and \ref{tab:llr_rmse} report the performance of KLNW with endpoint LLR (Sec.~2.8), and compare it with SVR.
We observe that combining endpoint LLR at inference can further improve $R^2$ and RMSE in several settings.

\begin{table*}[t]
  \centering
  \caption{$R^2$ for KLNW combined with endpoint local linear regression (LLR) at inference. Each cell reports (Train, Test).}
  \label{tab:llr_r2}
  \scriptsize
  \begin{tabular}{lcccccccccccc}
    \hline
    & \multicolumn{2}{c}{$S{=}100$} & \multicolumn{2}{c}{$S{=}200$} & \multicolumn{2}{c}{$S{=}500$}
    & \multicolumn{2}{c}{$S{=}1000$} & \multicolumn{2}{c}{$S{=}2000$} & \multicolumn{2}{c}{SVR} \\
    Dataset & Train & Test & Train & Test & Train & Test & Train & Test & Train & Test & Train & Test \\
    \hline
    bodyfat & 0.943 & 0.950 & 0.948 & 0.942 & 0.956 & 0.948 & 0.963 & 0.958 & 0.964 & 0.934 & 0.975 & 0.958 \\
    mg      & 0.745 & \textbf{0.629} & 0.779 & \textbf{0.645} & 0.794 & \textbf{0.648} & 0.796 & \textbf{0.654} & 0.798 & \textbf{0.652} & 0.853 & 0.562 \\
    energy  & 0.956 & 0.946 & 0.975 & 0.969 & 0.986 & \textbf{0.980} & 0.988 & \textbf{0.984} & 0.988 & \textbf{0.984} & 0.995 & 0.978 \\
    ccs     & 0.843 & 0.703 & 0.864 & 0.738 & 0.883 & 0.764 & 0.898 & 0.780 & 0.891 & 0.760 & 0.890 & 0.783 \\
    \hline
  \end{tabular}
\end{table*}

\begin{table*}[t]
  \centering
  \caption{RMSE for KLNW combined with endpoint local linear regression (LLR) at inference. Each cell reports (Train, Test).}
  \label{tab:llr_rmse}
  \scriptsize
  \begin{tabular}{lcccccccccccc}
    \hline
    & \multicolumn{2}{c}{$S{=}100$} & \multicolumn{2}{c}{$S{=}200$} & \multicolumn{2}{c}{$S{=}500$}
    & \multicolumn{2}{c}{$S{=}1000$} & \multicolumn{2}{c}{$S{=}2000$} & \multicolumn{2}{c}{SVR} \\
    Dataset & Train & Test & Train & Test & Train & Test & Train & Test & Train & Test & Train & Test \\
    \hline
    bodyfat & 0.004 & 0.004 & 0.004 & 0.005 & 0.004 & 0.005 & 0.004 & 0.004 & 0.004 & 0.005 & 0.003 & 0.004 \\
    mg      & 0.114 & \textbf{0.139} & 0.106 & \textbf{0.136} & 0.103 & \textbf{0.136} & 0.102 & \textbf{0.135} & 0.101 & \textbf{0.135} & 0.086 & 0.151 \\
    energy  & 2.095 & 2.449 & 1.565 & 1.863 & 1.189 & \textbf{1.482} & 1.073 & \textbf{1.337} & 1.073 & \textbf{1.326} & 0.713 & 1.542 \\
    ccs     & 6.428 & 9.268 & 5.990 & 8.704 & 5.551 & \textbf{7.778} & 5.175 & 7.962 & 5.370 & 8.320 & 5.392 & 7.916 \\
    \hline
  \end{tabular}
\end{table*}

\subsection{Learned spectral distribution}
\begin{figure}[t]
  \centering
  \includegraphics[width=0.95\linewidth]{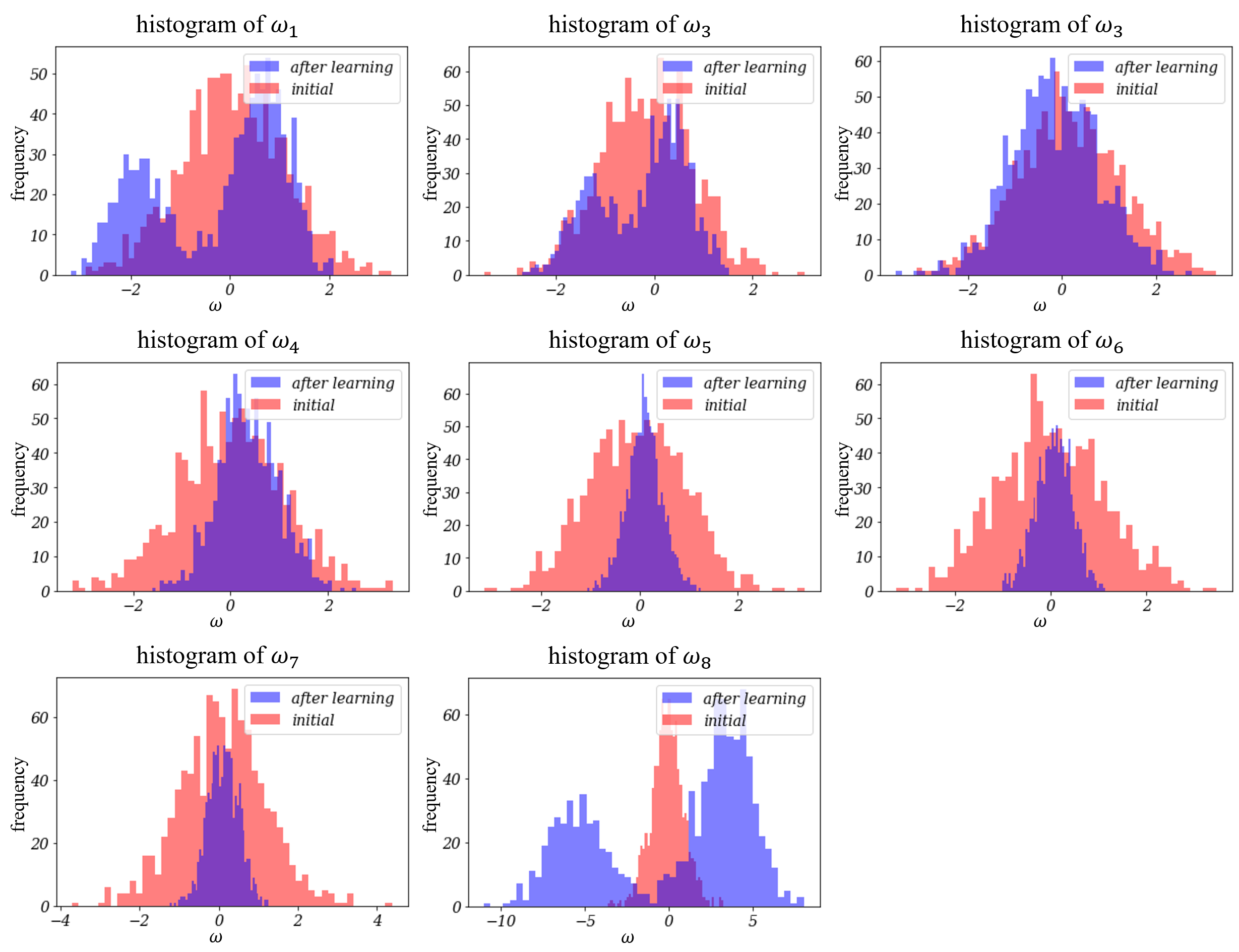}
  \caption{Histogram of sampled frequencies $\omega$ on the \textit{ccs} dataset ($S=1000$).
  The learned distribution differs from a single Gaussian.}
  \label{fig:hist}
\end{figure}

Finally, we visualized the distribution of sampled frequencies $\omega$ after training.
Figure~\ref{fig:hist} shows the histogram of $\omega$ samples on the \textit{ccs} dataset (with $S=1000$).
The learned distribution deviates from a single Gaussian, suggesting that the proposed approach can acquire
a data-adaptive spectral structure beyond fixed Gaussian RFF.

\section*{Discussion and Conclusion}

We proposed a QA-in-the-loop kernel learning framework for regression, in which a shift-invariant kernel is learned
through its spectral distribution.
Using Bochner's theorem, the kernel is expressed as an expectation over frequencies and approximated by a random Fourier features (RFF).
The spectral distribution is parameterized by an RBM (and its multi-layer extension), and discrete RBM states are
sampled by a quantum annealer (QA) and mapped to continuous frequencies via a Gaussian--Bernoulli transformation.
The resulting kernel is trained end-to-end by minimizing the leave-one-out Nadaraya--Watson (NW) mean squared error,
and we employed squared-kernel weights primarily to avoid near-zero NW denominators caused by cancellation in the finite-sample RFF approximation; as a secondary effect, this weighting increases the contrast of similarities.
In addition, we evaluated local linear regression using the same squared-kernel weights.

Experiments on multiple benchmark regression datasets showed that kernel learning decreases the training objective and induces clear changes in the kernel matrix, indicating that the similarity structure is adapted to the data.
Quantitatively, the learned kernel generally improves $R^2$ and RMSE compared with a fixed Gaussian-kernel NW baseline,
and increasing the number of random features at inference further enhances accuracy, consistent with reduced Monte Carlo approximation variance.
Moreover, local linear regression at inference can provide additional gains in settings where boundary bias or local non-uniformity affects NW regression.

Several directions remain for future work.
First, the relationship between QA-generated samples and the effective temperature/noise characteristics of hardware should be analyzed more systematically, and its impact on kernel quality should be quantified.
Second, scaling the spectral model beyond small RBMs---for example by improved embeddings, sparse or structured RBMs \cite{ohzeki2015},
or alternative generative models---is important for higher-dimensional problems.
Third, while we focused on regression, extending the same QA-driven spectral kernel learning to classification and uncertainty-aware regression (e.g., by coupling with probabilistic models \cite{rasmussen2006}) is promising.
Finally, a more detailed theoretical understanding of generalization under leave-one-out training with stochastic RFF kernels, including the role of squared-kernel weighting, would strengthen the foundation of QA-in-the-loop kernel learning.

In summary, our results demonstrate that QA-based sampling can be integrated into a complete kernel-learning pipeline and can yield practically useful, data-adaptive kernels for kernel regression.

\begin{acknowledgment}

We received financial support from the Cabinet Office programs for bridging the gap between R\&D and IDeal society (Society 5.0), Generating Economic and Social Value (BRIDGE), and the Cross-ministerial Strategic Innovation Promotion Program (SIP) (No.~23836436).

\end{acknowledgment}


\bibliographystyle{jpsj}
\bibliography{library}

\end{document}